\documentclass[prl,twocolumn,showpacs,preprintnumbers,amsmath,amssymb]{revtex4}

\usepackage{graphicx}
\usepackage{dcolumn}
\usepackage{bm}

\usepackage{color}

\usepackage{amsmath}	
\begin{document}


\title{
Photoinduced transition between conventional and topological insulators
in two-dimensional electronic systems
}

\author{
Jun-ichi Inoue 
}
\email{INOUE.Junichi@nims.go.jp}
\author{Akihiro Tanaka}
\affiliation{
National Institute for Materials Science,
1-1 Namiki, Tsukuba 305-0044, Japan}

\date{\today}

\begin{abstract}
Manipulating the topological properties of insulators,  
encoded in invariants such as the Chern number and its generalizations, 
is now a major issue 
for realizing novel 
charge/spin responses in electron systems.   
We propose that a simple optical means 
--subjecting to 
a driving laser field with circular polarization,   
can be fruitfully incorporated to this end. 
Taking as a prototypical example the two-band insulator 
 first considered by Haldane,  
we show how the electron system can be tuned through phases associated 
with different Chern numbers as the laser intensity is adiabatically swept;   
i.e., a photo-induced analog of the quantum Hall 
plateau transition. The implications of our findings includes 
the possibility of laser-tuning a conventional insulator into a quantum
 spin Hall system.  

\end{abstract}

\maketitle


The past few decades have seen the topology-related quantum aspects of
electron systems  
rise to the forefront of condensed matter physics\cite{Bohm}. 
The unifying concept behind these developments is the Berry phase, 
i.e., an extra phase which the wavefunction picks up 
as the system undergoes an adiabatic excursion. 
In a semiclassical 
description of Bloch electrons, the relevant excursion takes place in
momentum space, 
motivating the introduction of the Berry connection  
${\cal A}({\bf k})=i\langle v({\bf k})\vert\nabla_{\bf k}\vert v({\bf
k})\rangle$, where $v({\bf k})$ is the  
periodic part of the Bloch wavefunction; a contour integral of ${\cal
A}(\bf k)$ gives   
the Berry phase associated with that path. Also playing a crucial role is  
the Berry curvature $\Omega({\bf k})=\nabla_{\bf k}\times{\cal A}({\bf k})$,  
which 
adds an 
anomalous-velocity term to the standard equation of motion of an
electron\cite{Karplus}. 
The incorporation of these  
new entities has deeply enriched modern solid state physics with e.g., a
fully quantum mechanical  
theory of electron polarization\cite{King-Smith,Resta}, the notion of
adiabatic quantum pumping\cite{pumping}, an intrinsic mechanism of  
the anomalous Hall effect\cite{Onoda}, and schemes of generating a
dissipationless spin current\cite{Murakami}.     

Largely triggered by the discovery of the quantum spin Hall effect (QSHE), 
these activities have in more recent years culminated in 
the conception, classification, and an intensive search for  
new states of matter 
collectively referred to as {\it topological insulators} (TI). 
In its broadest sense, TIs are bulk insulators which nevertheless exhibit 
novel linear responses to external fields. These responses are dominated
by transport through  
quantized edge/surface channels, whose robustness is ``protected'' by a
set of topological invariants   
characteristic to that system. A well known example of the latter is the 
Chern number of a 2d system, ${\cal C}\equiv \frac{1}{2\pi}\int_{\rm
BZ}\frac{d^{2}k}{(2\pi)^{2}}\Omega({\bf k})$,  
where the integration is to be performed with respect to the first Brillouin zone. 
Introduced into electron physics by Thouless {\it et al} in their pioneering work on 
the integer quantum Hall effect (IQHE), this integer-valued quantity
basically measures  
the effective ``magnetic flux'' intrinsic to  
the Bloch electrons, thus giving rise to a quantized Hall 
conductivity $\sigma_{xy}=\sum_{n}{\cal C}_{n}\frac{e^{2}}{h}$, 
with the summation taken over all occupied bands\cite{Thouless}. 
Hence 
a {\it Chern insulator} (CI), i.e., a 2d TI 
with ${\cal C}_{\rm eff}\equiv\sum_{n}{\cal C}_{n}\ne 0$,   
realized 
in the absence of a magnetic field 
exhibits a {\it spontaneous quantized Hall effect}.  
States corresponding to different values of ${\cal C}_{\rm eff}$ are to
be regarded as belonging to  
distinct phases, as is the case with each $\sigma_{xy}$-plateau phase of
the IQHE.     
Haldane 
was the first to devise  
a minimal two-band model without a magnetic field which accommodates CI
phases\cite{Haldane}.   
Despite the artificial nature of its construction, this model 
bears immediate relevance to 
numerous physical systems (see later discussions). 
Furthermore the essence of QSHE is understood in terms of  
a superposition of two copies 
of Haldane models (one for each spin component) with 
a net time-reversal symmetry\cite{Mele,Qi}; recent work clarifies how  
this picture relates to a more generic characterization of QSHE 
employing ${\bf Z}_{2}$-valued invariants. 

From the perspective of material function engineering, 
a major shortcoming which plagues the search for TIs is that 
their defining topological invariants, such as ${\cal C}_{\rm eff}$,   
are usually material constants whose values cannot be varied freely
within a given sample.    
This is to be contrasted with the situation in IQHE\cite{Thouless}, where  
an adiabatic sweeping of the magnetic field strength drives the system from 
a normal (${\cal C}_{\rm eff}=0$) to a topological (${\cal C}_{\rm
eff}\ne 0$) insulator,  
and vice versa. In this Letter, we pose and answer in the affirmative the question: 
is there a generic approach allowing us to externally tune 
the topological invariant of a TI? 
 Taking up the most basic example of the Haldane model, 
we show below that an optical means  
--the application of a monochromatic driving laser field 
with circular polarization, 
can be invoked to obtain a {\it photo-induced normal-Chern insulator transition}; 
i.e., by slowly changing 
the driving laser amplitude (or the laser frequency), 
one can transport the electron system into different regimes 
of the ${\cal C}_{\rm eff}$ vs model parameter phase diagram. This scenario
naturally lends itself 
to more generic topological phase transitions as we later mention, and 
thereby suggests  a route to manipulating the topological properties of
various insulators\cite{Thonhauser}.

Let us then illustrate how the claimed result comes about. 
We begin by noting that our purpose requires us to 
carefully trace the low-energy sector of the electron system through its evolution,  
as the laser field intensity is adiabatically swept. 
This we accomplish by adopting a systematic method previously developed by 
one of the authors\cite{Inoue}. 
Aided by the powerful machinery of the Floquet theory, the 
procedure sketched below is applicable to the adiabatic dynamics of any tight-binding model, 
and was previously used to 
give a reliable prediction on how the electric polarization 
of a (normal) insulator 
renormalizes substantially when subjected to a 
slowly varying laser field\cite{Inoue}. 
While the polarization is also a prominent
Berry phase effect,   
it is worth stressing that the adiabatic flow of a {\it global}
topological quantity such as ${\cal C}_{\rm eff}$  
is considerably harder to foresee. 
Consider then the Hamiltonian for the Haldane model $H_{H}$ coupled to a
circularly polarized driving laser field  
$H(t)=H_{H}+H'(t)$, where
\begin{align}
& H_{H}=
t_{1}\sum_{i}\sum_{{\vec{r}}\in
 A}\left[a^{\dagger}(\vec{r})b(\vec{r}+\vec{d}_{i})
+{\rm H. C. }\right] \nonumber \\
&+t_{2}\sum_{j}\Bigg[
\sum_{\vec{r}\in A}e^{i\phi}a^{\dagger}(\vec{r})a(\vec{r}+\vec{\ell}_{j})
+\sum_{\vec{r}\in
 B}e^{-i\phi}b^{\dagger}(\vec{r})b(\vec{r}+\vec{\ell}_{j})
\nonumber \\
&\qquad+{\rm H. C. }\bigg]
+
\Delta\sum_{\vec{r}\in A}a^{\dagger}(\vec{r})a(\vec{r})
-\Delta\sum_{\vec{r}\in B}b^{\dagger}(\vec{r})b(\vec{r}),
\\
&H'(t)=e\vec{E}(t)\cdot
\Bigg[
\sum_{\vec{r}\in A}\vec{r}a^{\dagger}(\vec{r})a(\vec{r})+\sum_{\vec{r}\in
 B}\vec{r}b^{\dagger}(\vec{r})b(\vec{r})
\Bigg].
\end{align}
The electrons reside on a 2d honeycomb lattice 
comprising two triangular 
sublattices which we denote as $A$ and $B$; 
we associate with each the electron creation operators $a^{\dagger}$ and
$b^{\dagger}$.   
The vector $\vec{d}_{i}$ ($\vec{\ell}_{j}$) connects nearest (next-nearest) 
neighbors [see Fig.\,1]. 
The system is at half-filling, and is thus a band insulator. 
The staggered potential $\Delta$ and the phase $\phi$ are 
each responsible for breaking inversion and time-reversal symmetries, 
which causes the lower-band Chern number to assume the set of possible
values ${\cal C}=0, \pm1$\cite{Haldane}.   
\begin{figure}
\includegraphics[width=5.5cm]{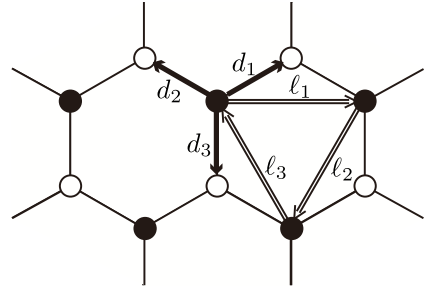}
\caption{The lattice structure considered in this study.  
Solid (open) dots 
{
 denote sites belonging to the} $A$ ($B$) sublattice.  
{
Electrons residing on the} $A$ ($B$) sublattice 
{
suffer} an on-site 
electron potential of $+\Delta $ $(-\Delta)$; this staggered nature 
breaks the inversion symmetry. 
The electron hopping energies along 
$\vec{d}_{i}$ and $\vec{\ell}_i$ are $t_{1}$ and $t_2 e^{i\phi}$, respectively.  
}
\label{lattice}
\end{figure}
Turning to the interaction term $H'$, 
we choose the wave vector of the laser field to be normal to the lattice plane. 
The driving laser field with a fixed frequency $\omega$ 
can be expressed as 
$\vec{E}(t)=E_{T}(\cos\omega t, 
\tau\sin\omega t)$, where $\tau=+1 (-1)$ 
corresponds to
the left (right)
circular polarization.  
Now let the 
time-dependence of the amplitude $E_T$ be weak enough to justify 
the use of the adiabatic approximation, 
which is always possible by preparing 
a suitable laser source.
Then the sinusoidally time dependent problem 
conveniently maps onto
a time-independent eigenvalue problem 
$\tilde{H}_{F}|\Phi\rangle
=\tilde{\varepsilon}|\Phi\rangle$ 
by virtue of the Floquet theory\cite{Shirley,Sambe};    
the eigenvalue $\tilde{\varepsilon}$ is 
the quasi-energy 
in the language of the latter framework. 
A straightforward calculation shows that the Floquet Hamiltonian $\tilde{H}_{F}$ 
consists of $2\times 2$ block submatrices, 
i.e., $\tilde{H}_{F}=\big(\tilde{H}^{(mn)}\big)$ 
where the 
entries for each block (in the Fourier-transformed bases, $a_{k}$
and $b_{k}$) are:  
\begin{align}
 \tilde{H}^{(mn)}_{11}&=(\Delta
 +m\hbar\omega)\delta_{mn}+t_2J_{N}(\sqrt{3}\lambda)\sum_{j=1}^{3}e^{i\tau\frac{2N(j-1)\pi}{3}}  
 \nonumber \\ 
&\times \bigg[
(-1)^Ne^{i(\vec{k}\cdot\vec{\ell}_{j}+\phi)}
+
e^{-i(\vec{k}\cdot\vec{\ell}_{j}+\phi)}
\bigg], 
\label{H11}
\\
\tilde{H}^{(mn)}_{12} &= t_1J_{N}(-\lambda)\sum_{j=1}^3
e^{-i\tau\frac{N(4j-3)\pi}{6}}e^{ i\vec{k}\cdot\vec{d}_{j}},
\label{H12}
\end{align}
with $J_{N}$ the Bessel function of 
the $N\equiv n-m$
--th order and
$\lambda\equiv eaE_{T}/\hbar\omega$ ($a$: lattice constant).   
The remaining matrix elements, $\tilde{H}^{(mn)}_{22}$  and $\tilde{H}^{(mn)}_{21}$,
are identical in form to 
Eqs. (\ref{H11}) and (\ref{H12}), respectively, except for the insertion
of -1 in front of
$\Delta$ and $\phi$ in the former, and $\lambda$ and $\vec{k}$ in
the latter.  
This Hamiltonian operates 
on wavefunctions defined in the space 
compositely spanned by the electronic degrees of freedom and photons
with energy $\hbar\omega$.   
The diagonal block of the Hamiltonian, $\tilde{H}^{(nn)}$,
 is the $n$-photon sector, i.e., the subspace with $n$ photons.

Now suppose that the condition $W \ll \hbar\omega \ll E_{G}$ is met,  
where $W$ is the electronic band width and $E_{G}=2\Delta$ the energy gap.
In this case 
the admixture of neighboring photon sectors is negligible, 
which enables us to concentrate on the zero-photon sector\cite{Eckardt},
$
 \tilde{H}^{(00)}
=\sum_{k}
\begin{pmatrix}
a^{\dagger}_{k}, b^{\dagger}_{k}
\end{pmatrix}
\tilde{{\cal H}}
\begin{pmatrix}
a_{k} \\
b_{k}
\end{pmatrix}
$, 
\begin{align}
 \tilde{{\cal H}}&
=2t_2J_{0}(\sqrt{3}\lambda)\cos\phi
\sum_{j}\cos(\vec{k}\cdot\vec{\ell}_{j}){\bf I} \nonumber\\
&+t_1J_{0}(\lambda)
\sum_{i}
\left\{
\cos(\vec{k}\cdot\vec{d}_{i})\sigma_x
-\sin(\vec{k}\cdot\vec{d}_{i})\sigma_y
\right\}\nonumber \\
&+\Bigm[
\Delta-2t_2J_0(\sqrt{3}\lambda)\sin\phi
 \sum_{j}\sin(\vec{k}\cdot\vec{\ell}_{j})\Bigm]\sigma_z, 
\end{align}
where ${\bf 1}$ is a $2\times2$ unit matrix and $\sigma_{x,y,z}$ are the
Pauli matrices.  
It is straightforward to see that
$\tilde{{\cal H}}$
is identical in form to the Haldane model, save for the fact 
that the hopping integrals $t_1$ and $t_2$ are modulated by factors involving 
Bessel functions. Hence, expanding $\tilde{H}^{(00)}$ around the two Dirac points 
$K_{\alpha=\pm}=(\alpha
4\pi/3\sqrt{3}a, 0)$,  
we finally arrive at our  
effective Hamiltonian
\begin{align}
& \tilde{\cal H}_{\alpha}=
-3t_{2}J_{0}(\sqrt{3}\lambda)\cos \phi\, {\bf 1}
+\frac{3a}{2}t_{1}J_{0}(\lambda)\left[
-\alpha k_{x}\sigma_{x}+k_{y}\sigma_{y}\right]\nonumber 
\\
&\qquad +\left[
\Delta-3\sqrt{3}\alpha t_{2}J_{0}(\sqrt{3}\lambda)\sin\phi
 \right]\sigma_{z},
\label{effective_H}
\end{align}
where wave numbers $k_{x,y}$ are 
now measured 
relative to the $K_{\pm}$
points.  
Note that the zero-photon sector is insensitive to the direction of the
circular polarization.  
It is an easy task to read off from the 
results of Ref.\onlinecite{Haldane} that the Chern number for our 
effective system described by Eq. (\ref{effective_H}) is given by
\begin{align}
 {\cal C}=&\frac{1}{2}\sum_{\alpha=\pm}
\alpha\,{\rm sgn}\left(\Delta+\alpha3\sqrt{3}t_{2}J_{0}(\sqrt{3}\lambda)\sin\phi\right).
\label{main_result}
\end{align}
This expression for 
{\it the photo-induced modification of the Chern number} 
represents a principal result of this Letter:  
in addition to the two variables $\phi$ and $\Delta/t_{2}$  
governing the topological property of the original Haldane model\cite{Haldane}, 
we now have a {\it tunable third variable} $\lambda$ at our disposal
which comes from the coupling to the laser field.  
The extended Chern number phase-diagram within our three-dimensional parameter space 
is depicted in Fig.\,2.
Points 
belonging to the interior of the ravioli sheet-like structure 
fall within one or another of the Chern insulator phases,   
with either ${\cal C}=+1$ or ${\cal C}=-1$. 
Elsewhere the system is a normal insulator with ${\cal C}=0$.   
The array of ${\cal C}=+1$ and ${\cal C}=-1$ phases constitute a
checkerboard-like pattern.  
By simply sweeping the laser
amplitude, one can traverse 
the phase diagram along the $\lambda$
direction and 
cross phase boundaries, thus inducing normal-Chern insulator 
transitions. 
We have confirmed 
that the gap closes and reopens in the vicinity of the phase transition
point, as it must,  
since ${\cal C}$ is robust to a continuous deformation of the energy
bands\cite{Thonhauser}. 
We note in passing that the circular polarization  
was essential in the above; a replacement with a linearly
polarized field would fail to yield a 
massive Dirac-fermion-like energy spectrum as in Eq.(\ref{effective_H})
\cite{Zhao2008}.

\begin{figure}
\includegraphics[width=8.5cm]{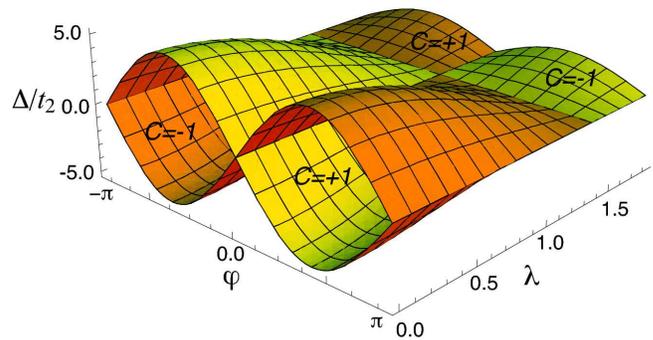}
\caption{(color online). The three-dimensional phase diagram of the normal and Chern
 insulators. (See text for details.)
}
\end{figure}

We now run a consistency check among several conditions which were
implicit above;
they will place some restrictions on the parameter values  
for which the transition is expected to be observable.
We first note that $|\Delta/t_2|<3\sqrt{3}$, 
which derives from the observation that $\lambda=0$ occupies the largest portion 
of the phase diagram with ${\cal C}\neq 0$. 
This needs to be consistent with the aforementioned condition for the
hopping energy,  
$W\sim t_1\ll\hbar\omega\ll\ E_{G}$. 
That, however is not straightforward, because on top of these 
 $|t_2J_0(\sqrt{3}\lambda)/t_1J_{0}(\lambda)|<1/3$ 
should also be satisfied in order to 
prevent the bands from overlapping\cite{Haldane}.   
To confront with this problem, we will relax the condition 
$W\ll\hbar\omega\ll E_{G}$ to $W<\hbar\omega< E_{G}$, which in turn will 
generally impose a restriction on the range of values, $[0,\Lambda]$, which the 
parameter $\lambda$ can assume without breaking down our
scheme.
(Note that when $W\ll\hbar\omega\ll E_{G}$ is satisfied, 
there are no such restrictions on $\lambda$.)  
We have evaluated this range by numerically checking  
the agreement between the quasi-energy bands of the Floquet Hamiltonian 
terminated at a certain order and the eigenenergy bands of
$\tilde{H}^{(00)}$.  
When $t_2/t_1=1/3$ and $\Delta/t_{1}=1.3$, 
we find that our scheme is valid 
for $\Lambda\simeq0.3$. We therefore confine our discussion to this
regime in subsequent discussions.  
The region where the normal-Chern insulator
transition occurs is 
displayed in Fig.\,3(a); 
in the filled-area, one can traverse between 
the interior and the exterior of the sheet structure
by changing $\lambda$ within 
the interval $0< \lambda \lesssim \Lambda$.

\begin{figure}
\includegraphics[width=8.cm]{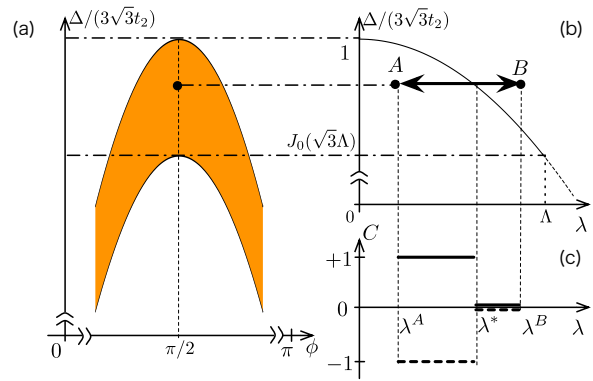}
\caption{(color online). (a) Partial section of the phase diagram 
{
for} $\lambda=0$. In the
 filled-area, one can traverse between 
the interior and exterior of the 
double-sheet structure by
changing $\lambda$ within the interval $0< \lambda \lesssim \Lambda$.  
(b) Section of the phase diagram 
{
for} $\phi=\pm\pi/2$.  
(c) Corresponding Chern number as a function of $\lambda\in[\lambda^A,
 \lambda^B]$ for $\phi=\pi/2$ (thick solid line) and $\phi=-\pi/2$
 (thick dashed line).  }	
\end{figure}

The normal-to-Chern insulator transition that we found 
can be viewed as a photo-induced analog of the 
inter-plateau transition in IQHE, 
wherein the laser field intensity replaces the role 
of the external magnetic field. (Recall that each IQHE Hall-plateau 
has a different value of ${\cal C}_{\rm eff}$.)
To strengthen this analogy it is instructive to 
consider sweeping the system between the two points 
 $A$ and $B$ in Fig.3(b), which
are the sections of Fig.2 at $\phi=\pm\pi/2$, and explicitly display  
how the Chern number change actually occurs.  
The path connecting the two points crosses the phase boundary at
 a critical point $\lambda=\lambda^{*}$, 
where a ``inter-plateau transition'' takes place as depicted in  
Fig.\,3 (c) in thick solid (dashed) lines.   
The field strength for the transition around the value of
$\lambda^{*}\sim {\cal O}(0.1)$ 
is estimated to be of the order of $\sim{\cal O}(10^{6})$[V/cm].
This is to be compared, e.g., with 
the corresponding estimate of $\sim {\cal O}(10^7)$ [V/cm] \cite{Dunlap} 
made in relation to {\it dynamic localization}, a 
prototypical example of optical manipulation of electron dynamics.
The above parallelism loosely conforms with earlier work ({\it not}  
straightforwardly applicable to our adiabatic process)   
suggesting that a circularly polarized light should affect electron motion in an
manner analogous to a magnetic field \cite{Ziel}.

Let us now size up what the implications of our strategy 
are for various physically motivated systems. 
(1) It has been shown\cite{Ohgushi} 
that the distribution of the Aharanov-Bohm-like flux inherent 
in the Haldane model (described by the phase $\phi$) can be mimicked 
by tight-binding electrons on a Kagome lattice moving in the background 
of a spin chirality ordering. This idea has lead to a detailed   
analysis on the anomalous Hall effect in several pyrochlore ferromagnets,
such as ${\rm Nd_{2}Mn_{2}O_{7}}$\cite{Taguchi}. 
Our method readily carries over to this system; indeed, in the
notations of Ref.\onlinecite{Ohgushi}  
we find that applying a circular polarized light to this particular
model yields ${\cal C}
=-{\rm sgn}\left(J_{0}(\lambda)\sin\phi\right)$
as the counterpart to Eq.(\ref{main_result}), from which we expect to
encounter the photo-induced transition  
${\cal C}
=-1\rightarrow {\cal C}
=+1$ as a function of laser
intensity. (2) It has also been proposed that the Haldane model can be
simulated by a fermionic cold-atom on an optical lattice\cite{cold_atom}. 
A simple estimation  
reveals that the energy scale of the laser fields which constitute the
lattice 
($\sim 1$ [eV])
and that of the  
additional sweeping field ($\sim 10^{-10}$ [eV] or larger) are easily
separated, which makes the implementation of the adiabatic scheme  
feasible. (3) The canonical Kane-Mele model\cite{Mele} for the QSHE, in
the absence of a Rashba term, reduces to  
two copies (up-spin and down-spin) of Haldane models. (The Luttinger
model for light/heavy hole bands in a semiconductor  
also reduces to the same model, provided we replace spins by a conserved
pseudospin quantum number\cite{Murakami2,Qi}.)  
The sum of the Chern numbers ${\cal C}_{\uparrow}$ and ${\cal C}_{\downarrow}$ for 
each spin component cancels in accord with time-reversal symmetry, 
while the difference ${\cal C}_{\uparrow}-{\cal C}_{\downarrow}$ needs not.   
The preceding generalizes trivially to this situation, which now leads
to a photo-induced normal to quantum spin Hall insulator (QSHI) transition.  
When the Rashba coupling is present, the ${\bf Z}_{2}$-invariant which takes over the role of ${\cal C}$ can be evaluated 
following Ref.\onlinecite{Fu-Kane}, using a matrix-valued generalization of the Berry connection. Here again we find that 
the system can be laser-tuned through a normal-QSHI  transition. 
Details on work along this line will be reported elsewhere.

To summarize, we have demonstrated how a photo-induced normal-Chern
transition occurs in the Haldane model.  
The coupling to a driving laser field with circular polarization
turns the transfer integrals into tunable variables which 
can result in a change in the Chern number.  
The strategy is reasonably general, and suggests an optics-based scheme
for manipulating the  
intrinsic topology of insulators.    

We thank M. Arai for helpful discussions. 
The authors were supported in part by Grant-in-Aid for Scientific Research 
(C) 20540386 and 22540340 from MEXT, Japan. 
%

\end{document}